\documentclass[prb,aps,epsfig,twocolumn,showpacs]{revtex4}
\usepackage{color}
\usepackage{epsfig,amssymb,amsmath,latexsym}
\usepackage{subfigure}
\usepackage{wrapfig}
\usepackage{float}
\usepackage{color}
\usepackage{bm,bbm,dsfont,ulem,nicefrac}
\usepackage[colorlinks=true,linktoc=page,linkcolor=magenta,citecolor=magenta]{hyperref}

\newcommand\bea{\begin{eqnarray}}
\newcommand\eea{\end{eqnarray}}
\newcommand\beq{\begin{equation}}  
\newcommand\eeq{\end{equation}}

\newcommand{\bib}{\bibitem}

\newcommand{\sq}[2][0]{
  \mbox{$\medmuskip=#1mu\displaystyle#2$}
  }
  
\begin{document}
\title{\bf{ Fingerprints of Majorana bound states in Aharonov Bohm  geometry  }}
\author{ Krashna Mohan Tripathi$^1$,
 Sourin Das$^2$ and Sumathi Rao$^1$}
\affiliation{$^1$ Harish-Chandra Research Institute, Chhatnag Road, 
Jhusi, Allahabad 211 019, India. \\
$^2$ Department of Physics and Astrophysics, University of Delhi,
Delhi - 110 007, India. }

\begin{abstract}
We study a  ring geometry, coupled to two normal metallic leads, which has a Majorana bound state (MBS) embedded in one of its arm and is threaded by Aharonov  Bohm ($\mathcal{A}\mathcal{B}$)  flux $\phi$. We show that  by varying the $\mathcal{AB}$ flux,  the two leads  go through resonance in an anti-correlated fashion while the resonance conductance is quantized to $2 e^2/h$. We further show that such anti-correlation is completely absent  when the MBS is replaced by  an  Andreev bound state (ABS). Hence this anti-correlation in conductance when studied as a function of $\phi$ provides a unique signature of the MBS which cannot be faked by an ABS.  We contrast the phase sensitivity of the MBS and ABS in terms of tunneling conductances. We argue that the relative phase between the   tunneling amplitude of the electrons and holes   from either  lead to the level (MBS  or ABS),  which is constrained to $0,\pi$ for the MBS and  unconstrained for the ABS,  is responsible for  this interesting 
contrast in the $\mathcal{AB}$ effect between the MBS and ABS.
 \end{abstract}
 
\maketitle
{\it{Introduction :-}} 
Zero energy Majorana bound states (MBS) which appear as end states of a 1-D p-wave superconductor have been attracting a lot of interest recently\cite{alicea,beenakker}, mainly due to their topological nature and relevance\cite{dassarma} in topological quantum computation. Although serious attempts  for confirming  the existence of the MBS have been made experimentally\cite{mourik,majoranaexpts},  their outcome remains controversial, and it is perhaps fair to say that there still has not been a definitive experiment to verify their existence. The primary reason for this  is that it  is not easy  to distinguish  Majorana modes from other spurious zero energy modes. This has also led to considerable  theoretical effort\cite{recentmajoranatheory} to look for clearly distinguishable robust signals of Majorana modes.

Many earlier theoretical studies have focussed on promising physical systems that support Majorana modes\cite{fukane,sau}.
Another focus\cite{oreg1,beenakker2} has been understanding and extending the proto-typical model that 
hosts Majorana modes, which
is the Kitaev model\cite{kitaev}. There have also been generalisations which yield more than 
one Majorana mode at each of the edges\cite{others, diptiman}, 
Floquet generation of Majorana modes\cite{others2,arijit}, etc. 

\begin{figure}[t]
\centering
\includegraphics[scale=0.2,width=0.35\textwidth]{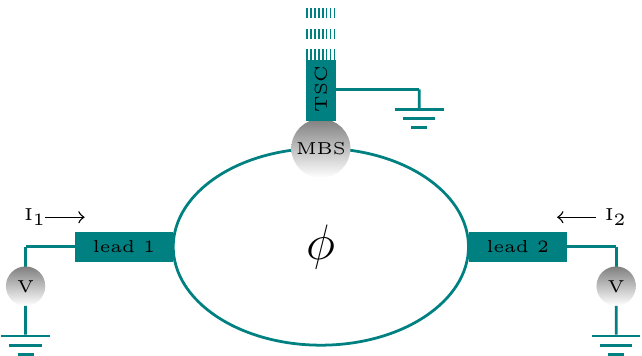}
\caption{ (color online) Schematic illustration  of the $\mathcal{AB}$ ring setup with two normal leads, at voltage $V$,  directly coupled to each other
as well as via a MBS/ABS hosted  at the edge of a grounded topological/non-topological  superconductor. 
}\label{fig1} 
\end{figure}

In this letter, we show that the Aharonov-Bohm ($\mathcal{AB}$) effect  in a ring geometry with a MBS embedded in one of its arm can provide a distinct signature which cannot be faked by an Andreev bound state(ABS).    Earlier attempts to use  $\mathcal{AB}$ flux  interferometers have been  in the
context of teleportation\cite{fu,jdsau} or non-local conductance or persistent currents\cite{benjamin}, but they involve the MBS at both ends of a wire.
Many  other recent proposals which discuss distinguishing signatures of the MBS  rely on quantum noise measurements\cite{oreg2} which are in general difficult to implement.
In contrast, we propose conductance measurements which can clearly distinguish the Majorana from a spurious zero mode.
Our proposed setup comprises of a two terminal ring geometry as shown in Fig.\ref{fig1},  with direct coupling between the leads as well as coupling via a MBS/ABS 
 hosted by a superconductor,  which is the  third lead  and which remains grounded for our proposal. We show that when  both the normal leads are equally biased with respect to the grounded superconductor, constructive resonance for one of the normal leads is always accompanied by a destructive anti-resonance on the other normal lead. As the conductance on each lead has flux periodicity of a flux quantum ($\phi_0=hc/e$), each normal lead  goes through a resonance and an anti-resonance as the  phase of the direct
 tunneling term, which is tunable by the 
 $\mathcal{AB}$ flux,  changes by $\phi_0$.  On the contrary,  when we replace the MBS by an ABS in the above described setup, we find that the current flowing through both the leads remains equal, irrespective of the variation of the $\mathcal{AB}$ flux. Hence the anti-correlation in current obtained as function of the flux  can be considered as a robust and direct signature of the MBS.

{\it Tunneling into the MBS:}
To begin with, we consider a model where a MBS is tunnel coupled to two normal leads. We will  later add a direct  tunneling term (with a complex phase) between the two leads to convert it into an effective Hamiltonian describing the  topological equivalent of a two path interferometer with an $\mathcal{AB}$ flux enclosed(see Fig.[\ref{fig1}]) where the 
$\mathcal{AB}$ flux is given by the phase of the complex tunneling amplitude. 
The Hamiltonian for the system in the absence of direct tunneling is given by 
\begin{eqnarray}
&H_0 = \underset{\alpha=1,2}{\sum}H_{\alpha} + H_{T} \quad {\rm with} \quad
H_{\alpha} = \underset{k}{\sum}\epsilon_{k}c^{\dagger}_{\alpha k}c_{\alpha k}  \nonumber \\
&{\rm and} \quad H_{T} = i \gamma \underset{\alpha,k}{\sum} ( u_{\alpha} c_{\alpha k} + u_{\alpha}^{\ast} c^{\dagger}_{\alpha k} )  \label{maj}
\end{eqnarray}
where the  index $\alpha =1,2$ runs over the two leads and $c_{\alpha k}$ corresponds to the electron operator in lead-$\alpha$. 
 $u_\alpha$ and $u_\alpha^*$ denote the complex amplitudes (matrix elements)  for the coupling of the MBS to the
electron and hole operators on the  two leads. 
From general considerations of hermiticity of the Majorana operator, 
the tunneling can only be to an `equal\rq{} linear combination of electron and hole annihilation operators on the normal lead given by  $c_{\alpha k}$ and $c_{\alpha k}^*$,
and by adjusting the phases of the basis states in the leads, we can choose the tunneling amplitudes $u_\alpha$ to be real.
Now it is straight forward to obtain the scattering matrix at zero energy for this  problem using the Weidenmuller formula\cite{weidenmuller} given by 
\beq
S = 1 - 2 \frac{u^{\dagger} u}{(u u^{\dagger})}=\begin{pmatrix}
   s^{ee} & s^{eh}\\
   s^{he} & s^{hh}\\
   \end{pmatrix}, ~  
u = \sqrt{\nu} 
\begin{pmatrix} 
u_{1} & u_{2} & u_{1}&  u_{2} \\
  \end{pmatrix} 
  \eeq
%
where the $s_{ee}$, etc, are $2\times 2$ matrices in the basis of the two leads and the full $S$-matrix has been written in the particle-hole basis.
Here,  $\nu$ is the density of states in the leads. Following Ref.\onlinecite{datta,nillson},  the current and noise at zero temperature and for 
a finite  bias on both leads can be obtained.  We give the general expression for arbitrary voltages on the two leads in the supplementary section
and confine ourselves to
vanishingly small but equal bias on both leads here, where this reduces 
\begin{eqnarray} 
&I_{i} = \frac{e^{2}}{h} \, V \, \frac{2u_{i}^{2} }{U},\quad
P_{11} = \frac{e^{3}}{h} \, (2V) \, \frac{u_{1}^{2} u_{2}^{2}}{U^{2}}=P_{22} \nonumber \\
&{\rm and}  ~~~P_{12} = - \frac{e^{3}}{h} \, (2V) \, \frac{u_{1}^{2} u_{2}^{2}}{U^{2}} = P_{21}
\end{eqnarray}
where $U=u_1^2+u_2^2$.
An important point to note here is the fact that sum of the conductances on both leads is fixed at  $I_1/V+I_2/V=2e^2/h$. In fact,  it remains quantized at  this value for any number of normal leads\cite{futpaper} tunnel coupled to the Majorana. This implies that the increase in conductance in a given lead has to be  compensated by a decrease in the other leads.  This feature is unique to the MBS and  is completely absent  for the  ABS.  
We will show later that this anti-correlation in the conductance between the two leads can be tuned via the flux  in an $\mathcal{AB}$ type set up. 
Further, we note that the quantization of the conductance actually implies a sum rule not just for the average current but for the sum of the current 
operator itself given by $\hat {I_1} +\hat{ I_2}= 2e^2/h~V$ at zero energy. This immediately implies that the fluctuation in the currents are strongly constrained, leading to the sum rule for the noise $\sum_{ij} P_{ij} =0$. Once  Fermi statistics is taken into account  in addition to the quantization of total conductance, this 
automatically implies that the auto-correlated noise is positive and the cross-correlated noise is  negative definite ($P_{11}=P_{22} = -P_{12} = -P_{21}$).
The issue of the cross-correlated noise being negative definite has has been addressed earlier in the context of the MBS\cite{lee} but its origin in the current sum rule has not been explicitly mentioned till now. 

{\it MBS and the $\mathcal{AB}$ set up:}
 Now we add  a direct  tunneling term (not via the MBS) between the two normal leads. This   converts the original Hamiltonian in Eq.\ref{maj}  into  a
 Hamiltonian that describes the topological equivalent of a two path interferometer with  $\mathcal{AB}$ flux enclosed(see Fig.[\ref{fig1}]), 
 where the $\mathcal{AB}$ flux is just given by the phase of the direct tunneling amplitude.  The phase freedom of the basis states
 on the leads has been used to make the tunneling to the Majorana mode real. Hence, the direct tunneling amplitude between the two
 leads will be complex in general. We choose the gauge of the $\mathcal{AB}$ gauge field to identify this phase with the $\mathcal{AB}$ flux.
 and write the  total Hamiltonian for the model  as 
\begin{eqnarray}
 H= H_0 + {\tau_0}\, e^{i \frac{2 \pi \phi}{\phi_0}} \, \psi^{\dagger}_{1} \, \psi_{2} + h.c. 
 \label{Htotal}
 \end{eqnarray}
where $\phi$ has the interpretation of the  $\mathcal{AB}$ flux  enclosed and $\tau_0$ is a real number representing the amplitude of direct tunneling between the leads. To find the scattering matrix for the $\mathcal{AB}$ setup,  we need to extend the standard form of the Weidenmuller formula to include the effects of the direct tunneling term which is shown explicitly in the supplementary section.
This gives us the scattering matrix
  \begin{eqnarray}
 S(E) &=& S_{N} - i \pi \nu (1 + i \pi \nu T)^{-1} W^{\dagger} \times \\ &&[E + i \pi \nu W (1 + i \pi \nu T)^{-1} W^{\dagger}]^{-1} 
 W (1 + S_{N}) \nonumber \\
 \label{S-matrix}
\rm{where} \quad  S_{N} &=& (1 + i \pi \nu T)^{-1}(1 - i \pi \nu T) .
 \end{eqnarray}
Note that in the absence of the direct tunneling term between the wires given by the $T$ matrix defined as
\bea
T &=& \begin{pmatrix}
        (\tau+\tau^{\dag}) & 0 \\
                 0   &  -(\tau+\tau^{\dag})^{\ast}\\
       \end{pmatrix} ~,
       \eea 
 the $S$-matrix reduces to the usual Weidenmuller formula
\beq
 S(E) =1 - 2 i \pi \nu W^{\dagger} [E + i \pi \nu W W^{\dagger}]^{-1} W \text{ for } T = 0.
 \eeq
It is now straight forward to obtain the current and the noise from the scattering matrix obtained by applying Eq.[\ref{S-matrix}] to  the Hamiltonian for the  $\mathcal{AB}$
 set up given in Eq.[\ref{Htotal}]. We find that 
\begin{flalign} 
& \tilde{I}_{1} = \frac{e^{2}}{h} V ~ \frac{2|\tilde{u}_{1}|^{2}}{\tilde{U}} , ~~
 \tilde{I}_{2} = \frac{e^{2}}{h}  V ~ \frac{2|\tilde{u}_{2}|^{2}}{\tilde{U}} \nonumber \\
 &\tilde{P}_{12} = -   \frac{e^{3}}{h} (2V)  ~  \frac{ |\tilde{u}_{1}|^{2} |\tilde{u}_{2}|^{2}}{\tilde{U}^{2}}   
               =  \tilde{P}_{21} = -  \tilde{P}_{11} = -  \tilde{P}_{22} \nonumber \\
&   \sq{       \tilde{u}_{1}= \sq{u_{1} + i  \pi \nu \tau_{0}   e^{- i \frac{2 \pi \phi}{\phi_0}}   u_{2}, \tilde{u}_{2} = u_{2} + i \pi \nu \tau_{0}  e^{i \frac{2 \pi \phi}{\phi_0}}  u_{1}}} \label{tun}. 
\end{flalign}  
Here $\tilde{U} =|\tilde{u}_{1}|^{2} + |\tilde{u}_{2}|^{2}$. Firstly we note that the sum of conductance $\tilde{I_1}/V+\tilde{I_2}/V=2e^2/h$ is still quantized and is independent of $\phi$ 
while the difference oscillates with the  period $\phi_0$. Due to the sum rule, the resonance  condition for the lead-1(2) is given by  $\tilde{I}_{2}=0$ ($\tilde{I}_{1}=0$). This is equivalent to having $\tilde{u}_{2}=0$ ($\tilde{u}_{1}=0$) respectively. Hence it is easy to see that the resonance condition, corresponding to having a conductance of $2 e^2/h$ on a given lead, depends on both the amplitude $\tau_0$ and  the phase  $  2 \pi \phi/\phi_0$. 
%

In Figs. 2a and 2b,  we show the conductances $G_i$  and the auto and cross-correlated noise $P_{ii}$ and $P_{ij}$ computed\cite{datta}
 (see supplementary section for more
details) as a function of the flux at finite temperature and finite voltage. Note that  the anti-correlation remains valid even at finite temperature and voltages.  However, the sum of the conductances is no longer quantised to be $2e^2/h$. Instead, it get multiplied by  $F(V,T)$, a decreasing function of $V$ and $T$. 
The exponential fall-off of the current, and the amplitude of the cross-correlations  as the bias is increased, at different temperatures,
 is also shown in Figs. 3a and 3b.

{\it Tunneling to ABS and phase sensitivity even in the absence of direct tunneling:-} Here we first consider an ABS which is tunnel coupled to two normal leads and later add a direct  tunneling term (with a complex phase) between the two leads to convert the system to an $\mathcal{AB}$ type set up. The Hamiltonian for the system in the absence of direct tunneling is given by Eq.\ref{maj}, except that now the tunneling term is replaced by 
\beq
H_{T} = a^{\dagger} \underset{\alpha,k}{\sum} ( t_{\alpha} c_{\alpha k} + v_{\alpha}^{\ast} c^{\dagger}_{\alpha k} ) + h.c. , 
\eeq
where the ABS is denoted by the resonant level creation operator $a^\dagger$ and the tunneling amplitudes to the electron
and hole states on the leads are given by $t_\alpha$ and $v_\alpha^*$ respectively. Note that there are no terms in the Hamiltonian for the  ABS itself as it is  at zero energy. (A toy lattice model for the Andreev bound state leading to the above coupling has been explicitly shown in the supplementary section.)
Our main aim is to contrast the results now with the earlier results  where the coupling was to a MBS. The scattering matrix can be obtained as before through the Weidenmuller formula and we have
\begin{eqnarray}
S &=& 1 - 2 W^{\dagger}(W W^{\dagger})^{-1}W  \quad {\rm where}  \nonumber \\
W &=& 
\begin{pmatrix}
 t_{1} & t_{2} & v^{\ast}_{1} & v_2^{\ast} \\
 -v_{1} & -v_{2}  & -t^{\ast}_{1}& -t^{\ast}_{2} \\
 \end{pmatrix}  \nonumber \\
\end{eqnarray}
Note  that the matrix  $W W^{\dagger}$ is singular for $t_{1}=v_{1}$ and  $t_{2}=v_{2}$. Hence, it is not possible to obtain a comparison between the ABS and the MBS  simply by `setting' $a=\gamma$ and $t=v$ directly in the scattering matrix. 

To understand this better, let us consider the single lead case, which can simply be written as 
\begin{eqnarray}
H_{T} &=& \underset{k}{\sum} \begin{pmatrix}      
      a^{\dagger} & a \\  
      \end{pmatrix}     
      W
      \begin{pmatrix}
    c_{1 k}\\
    c^{\dagger}_{1 k}\\
    \end{pmatrix}  \text{with}~~ 
W = 
\begin{pmatrix}
 t_{1} & v^{\ast}_{1} \\
- v_{1} & - t^{\ast}_{1} \\
 \end{pmatrix}  \nonumber \\ 
WW^{\dagger} &=&  \begin{pmatrix}
 |t_{1}|^{2} + |v_{1}|^{2}  & -2 t_{1} v^{\ast}_{1} \\
 -2 v_{1} t^{\ast}_{1} & |t_{1}|^{2} + |v_{1}|^{2} \\
 \end{pmatrix}  \nonumber \\ 
\rm{and}~~&&Det(WW^{\dagger}) = (|t_{1}|^{2} - |v_{1}|^{2})^{2}  
 \end{eqnarray}
We can now explicitly write this Hamiltonian in terms of 2 Majorana modes by changing to the Majorana basis -
$
a = (\gamma + i\tilde{\gamma})/{2}, 
a^{\dagger} = (\gamma - i\tilde{\gamma})/{2}$
so that the Hamiltonian can be rewritten as 
\begin{eqnarray}
H_{T} &=& \underset{k}{\sum} \begin{pmatrix}      
     i\gamma & i\tilde{\gamma} \\ 
      \end{pmatrix}     
      \tilde{W} 
      \begin{pmatrix}
    c_{1 k}\\
    c^{\dagger}_{1 k}\\
    \end{pmatrix} \nonumber \\  {\rm with} \quad
\tilde{W}  &=& 
\begin{pmatrix}
 -\frac{i}{2}(t_{1}-v_{1}) & \frac{i}{2}(t^{\ast}_{1}-v^{\ast}_{1}) \\
 -\frac{1}{2}(t_{1}+v_{1}) & -\frac{1}{2}(t^{\ast}_{1}+v^{\ast}_{1}) \\
 \end{pmatrix}  
 \end{eqnarray}
As can seen from here, either at  $t_{1} =v_{1}$ or at $t_1=-v_1$, one of the Majorana modes disappear from the Hamiltonian and 
the coupling matrix couples the lead only to a single MBS. This observation provides us with  a physical picture for describing the basic difference between a MBS and an accidental zero energy ABS. An ABS tunnel coupled to a lead  corresponds to having a simultaneous tunnel coupling of the lead with a pair of Majorana bound states. 
  This is the main reason for the  ABS and the MBS behaving differently when embedded in an $\mathcal{AB}$  ring.

The current and noise correlations for a vanishing small but equal voltage $V$ applied to  the two leads at zero temperature are given by
\begin{eqnarray}
   I_{1} &=& I_{2}  
         = \frac{e^{2}}{h} \, V \,  \frac{8|(t_{1}v_{2} - t_{2}v_{1})|^{2}}{D} \nonumber \\
      P_{11} &=& \frac{e^{3}}{h} \, V \sq{  \frac{8(|t_{1}|^{2} + |t_{2}|^{2} - |v_{1}|^{2} - |v_{2}|^{2})^{2}|(t_{1}v_{2} - t_{2}v_{1})|^{2}}{D^{2}}} \nonumber \\
       &=&    P_{22} = P_{12} = P_{21}  \quad {\rm where} \end{eqnarray}
\beq       
 \sq{
   D  =  (|t_{1}|^{2} + |t_{2}|^{2} + |v_{1}|^{2} + |v_{2}|^{2})^{2} - 4 |(t_{1}v_{1}^{\ast} + t_{2}v_{2}^{\ast})|^{2}}
 \eeq
Note that  here, the current on the two leads are equal, and there is  no constraint on the total conductance. 
This can be understood as follows.
The presence of a single MBS  creates an anti-correlation in the current between leads, but 
the second Majorana in the ABS, which is the time reversed partner of the first one, compensates for the first Majorana and 
eliminates  the anti-correlation,  making it completely symmetric. The conductance in each of the two leads can be tuned to its resonant 
value of $2 e^2/h$ when the various amplitudes satisfy the condition $|t_{1}|^{2} + |t_{2}|^{2} =|v_{1}|^{2} + |v_{2}|^{2}$ 
which can straight forwardly read off from the expression for the  noise. Hence the resonant value for the sum of the conductance of the 
two leads for the ABS is $4 e^2/h$ while it is $2 e^2/h$ for the MBS. Hence any observation of  total conductance exceeding $2 e^2/h$ can rule out 
the presence of MBS.

\begin{figure}[t]
\centering
\begin{subfigure}
\centering
\includegraphics[scale=0.2,width=0.46\textwidth]{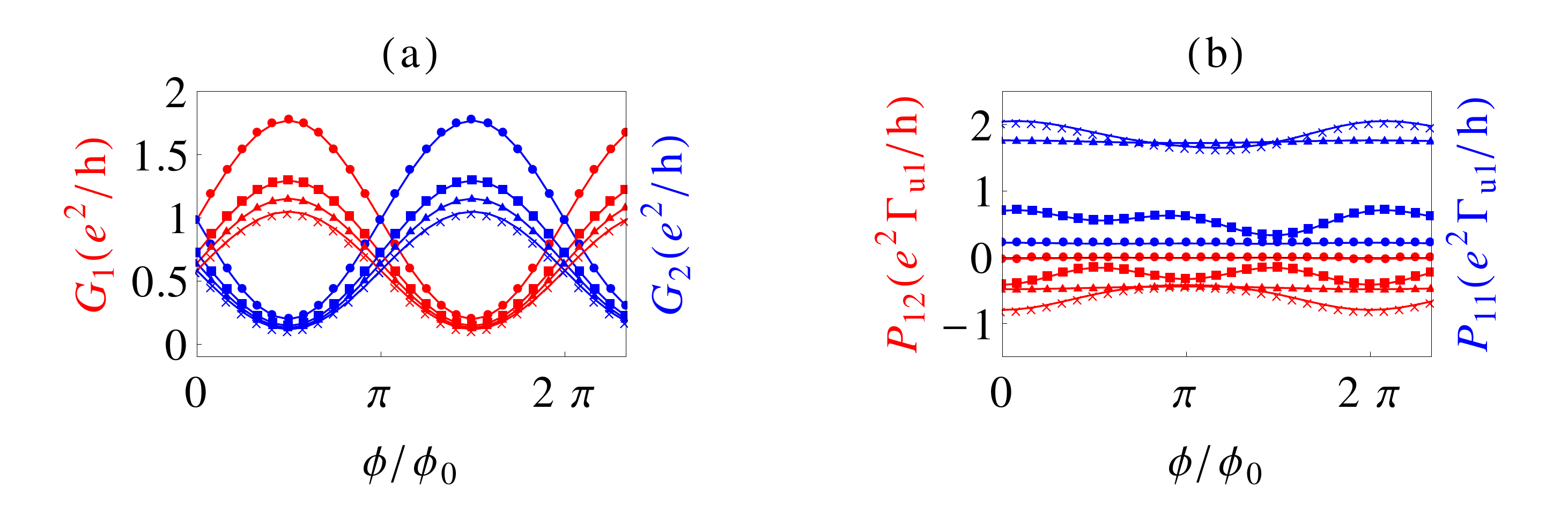}
\end{subfigure}
\begin{subfigure} 
\centering
\includegraphics[scale=0.2,width=0.46\textwidth]{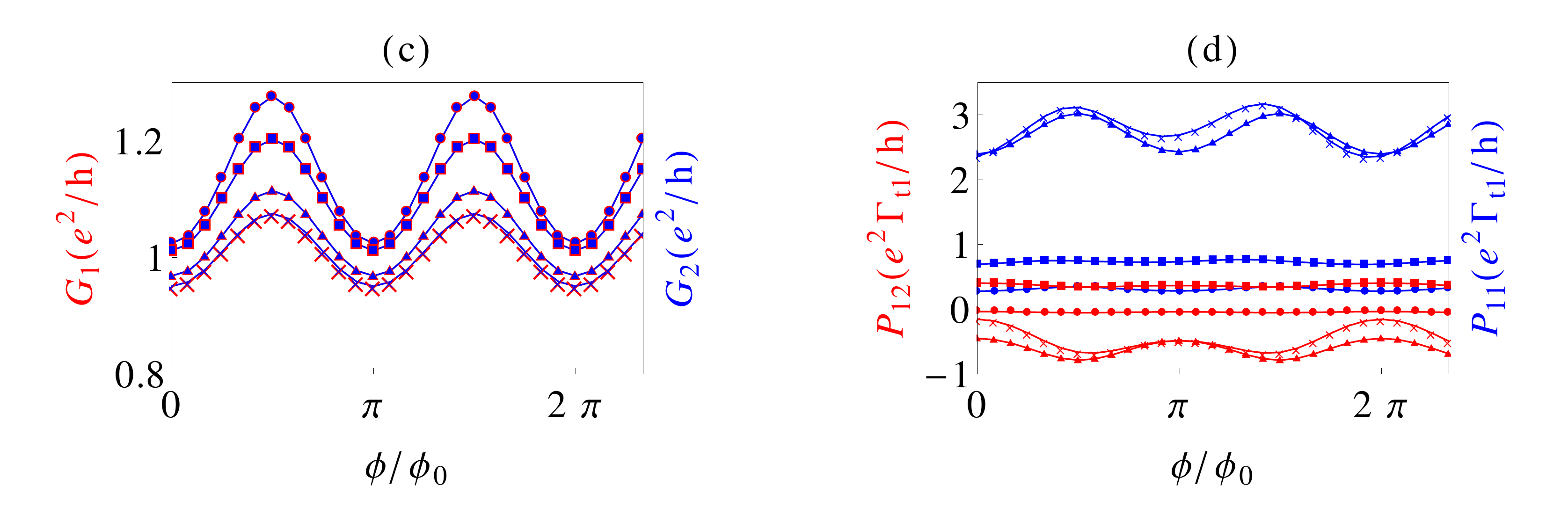}
\end{subfigure}



\caption{ (color online) {(a) and (b) ((c) and (d)) give the conductances on the two wires (red and blue)
and noise (red for cross-correlation and blue for auto-correlation)  for the $\mathcal{AB}$ interferometer with 
a MBS (ABS) on one of
its arms as a function of the flux at various  temperatures and voltages.  The anti-correlation of the currents in the MBS case and the 
positive correlation of the currents for the ABS case  as a function of the flux  survives even at finite temperatures and voltages.
Note also that the value of the total conductance $G_1+G_2$ is quantised to $2e^2/h$ for the MBS case, whereas it is non-universal and
can even go
above that value for the ABS case.
 The values of the temperature and voltage chosen are
(i) circles - $\mu=0.1\Gamma_{u1}$, $k_BT=0.1\Gamma_{u1}$ (ii) triangles - $\mu=0.1\Gamma_{u1}$, $k_BT=\Gamma_{u1}$
(i) squares - $\mu=\Gamma_{u1}$, $k_BT=0.1\Gamma_{u1}$ (ii) crosses - $\mu=\Gamma_{u1}$, $k_BT=\Gamma_{u1}$. 
 The  tunneling amplitude parameters are chosen to be $u_1=u_2=t_1=t_2=\frac{v_1}{2}=-\frac{v_2}{2}=\frac{\Gamma_{u1}}{\sqrt{2\pi\nu}}, \tau_0=
 \frac{\Gamma_{u1}}{2\pi\nu}$ with $\Gamma_{u1}(=2\pi\nu |u_1|^2) = \Gamma_{t1}(=2\pi\nu |t_1|^2) =1$.
}}
\label{fig2} 
\end{figure}

\begin{figure}[t]
\centering
\begin{subfigure}
\centering
\includegraphics[scale=0.2,width=0.46\textwidth]{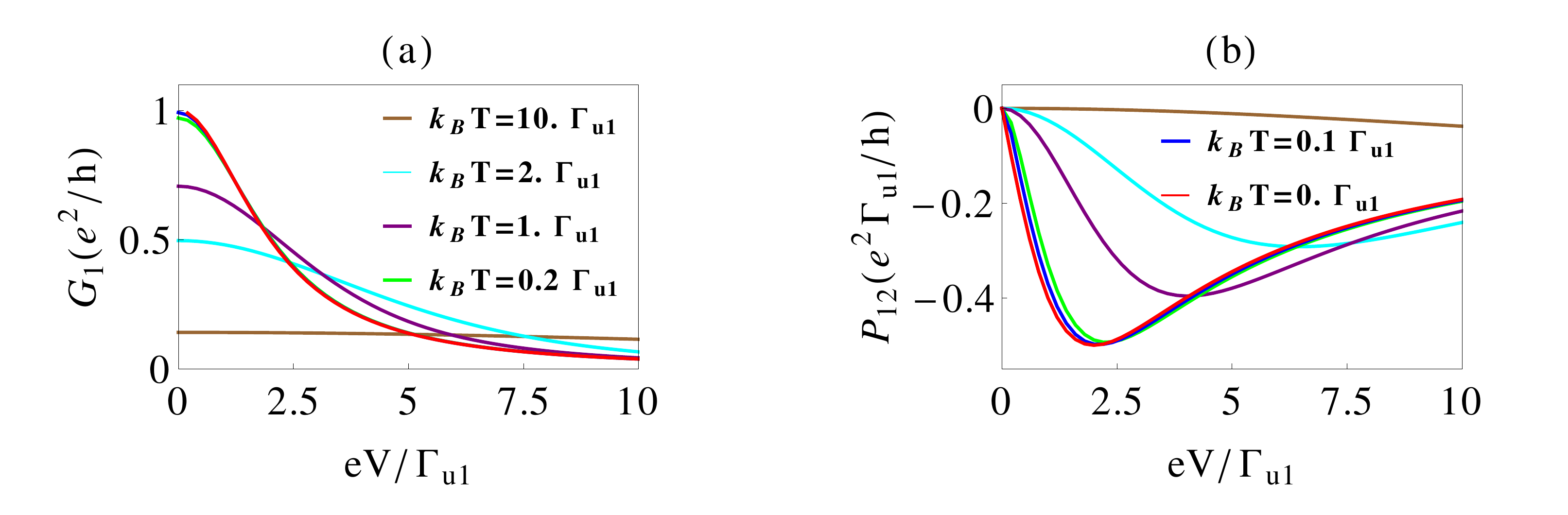}
\end{subfigure}
\begin{subfigure} 
\centering
\includegraphics[scale=0.2,width=0.46\textwidth]{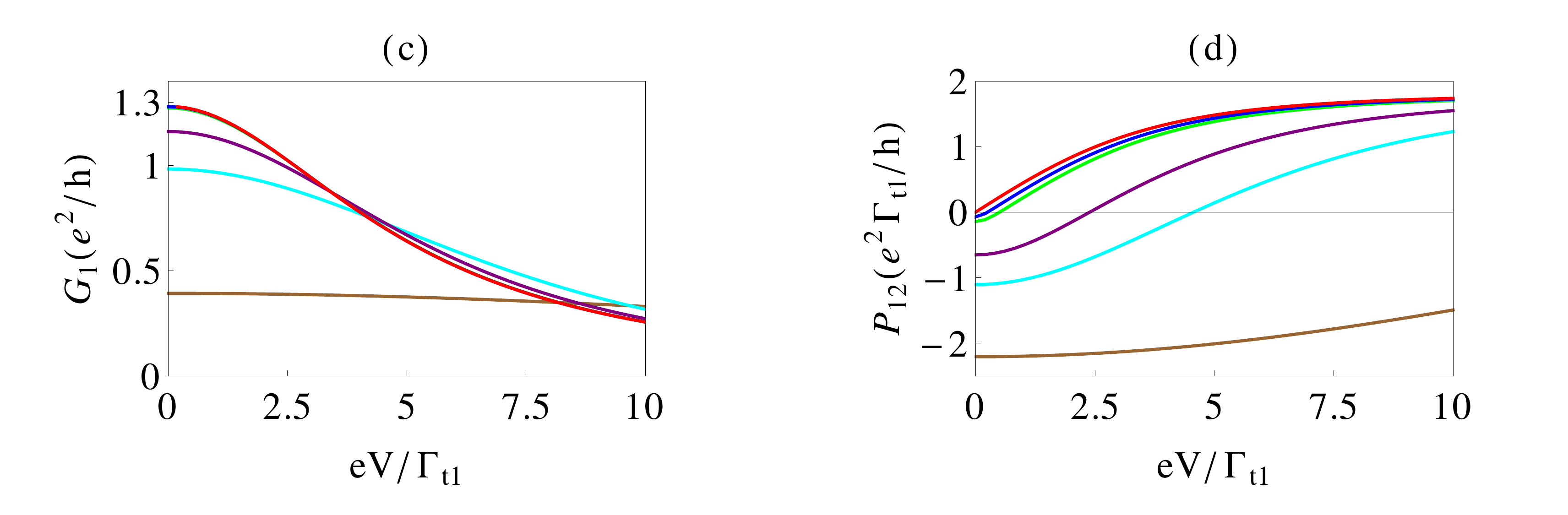}
\end{subfigure}
\caption{ (color online) {(a) ((c)) shows the exponential fall-off of the differential conductance on one of the wires 
as a function of the voltage for different temperatures for the MBS(ABS) case. For the MBS case, the zero bias peak
is quantised to $e^2/h$  (for a single wire) at zero temperature and then reduces to a non-universal value at finite
temperatures. For the ABS case, the zero bias peak even at zero temperature is non-universal. (b) ((d)) shows
the cross-correlations of the current through the two leads as a function of the voltage at various temperatures.
(The legend for the different colours is given in  (a) and (b) and is the same in all the figures.)
For the MBS case, the cross-correlations are always negative, whereas they can be positive or negative for the ABS case.
 The  tunneling amplitude parameters are chosen to be $u_1=u_2=t_1=t_2=\frac{v_1}{2}=-\frac{v_2}{2}=\frac{\Gamma_{u1}}{\sqrt{2\pi\nu}}$,  with $\Gamma_{u1}(=2\pi\nu |u_1|^2) = \Gamma_{t1}(=2\pi\nu |t_1|^2) =1$.
}}
\label{fig2} 
\end{figure}

We also note that the noise correlations on the two leads are positive, since the conductances are equal.
In fact, if one could tune the relative phase between the electron and hole terms
- i.e., choose $t_\alpha = |t_\alpha|e^{i\theta_{1\alpha}}$  and $v_\alpha= |v_\alpha|e^{i\theta_{2\alpha}}$, and tune 
the various $\theta_{i\alpha}$, not only the noise, even the conductance
would show oscillations as a function of any of the $\theta_{i\alpha}$. However, the  Majorana mode  only couples 
either to the case where the phases of the couplings to the electron and hole terms are the same
(the case chosen in the previous section), or when the coupling is to the other Majorana mode, where the phases 
differ by $e^{\pm i \pi} = -1$.  This  phase rigidity of the couplings to the electrons and holes in the leads is again a
feature of the MBS which is not shared by an accidental zero energy ABS. Hence, if this phase can be varied in a desirable 
fashion, it  can provide a distinguishing feature between a MBS and an accidental zero energy ABS. But in general 
this is not possible;  hence addition of the direct tunneling path with an enclosed flux  discussed in this letter provides a 
minimal set up for accessing the above described difference between the ABS and the MBS. 

Finally we also include direct tunneling between the leads to study the $\mathcal{AB}$ set up for  the ABS. However, since there 
are many parameters to vary, the results are highly dependent on the parameters chosen and the results for some typical values
are shown in Figs. 2c and 2d and Figs. 3c and 3d.  The only general feature that we 
see is that the conductance on the two leads remain identical, (but the cross-correlations can be positive or negative), irrespective of details 
of the $\mathcal{AB}$ flux, hence maintaining its contrast to the MBS case. This complete our study of $\mathcal{AB}$ set up for the ABS.

{\it Discussion and conclusion:-}
In this paper, we have attempted to distinguish between signatures of a MBS  and  an accidental zero
energy ABS by studying the conductance and noise correlations of two  leads in a two path interferometry 
setup with a superconductor (giving rise to a MBS or an ABS depending on whether or not
it is topological)  embedded in one of its arms.  By changing the phase of the direct tunneling between
the leads (equivalent to the $\mathcal{AB}$ phase), we find that the conductances in the two leads are perfectly
anti-correlated for the MBS case, with their sum quantized to be $2e^2/h$. Furthermore, the phase
of the direct tunneling can be tuned to give rise to a resonance in one of the leads, which is necessarily
accompanied by an anti-resonance in the other lead. This feature is completely absent
for the ABS and hence, can be used a strong fingerprint for the existence of a MBS.
We have also computed the noise correlations for both the MBS and the ABS, and attribute  the negative
cross-correlations in the MBS case to the strong correlation in the conductances on the two leads, 
coupled by the fermionic statistics of the MBS.  We point out  that for the coupling to the MBS, 
the phases between the electron and hole processes can only be either +1 or $-1$ (phase rigidity), whereas they 
can have an arbitrary phase for an ABS. This fact leads to distinguishing features in the transport across an $\mathcal{AB}$
set up. 
So the bottomline is that the distinction between  MBS and ABS is achieved via just conductance measurements alone.
In an $\mathcal{AB}$  ring geometry, the conductances in the two leads can be tuned by  the flux and the anti-correlation of the
conductances in the two leads is a strong fingerprint for the MBS. 

{\it Note added :-}  While writing up this work, the following papers [\onlinecite{fazio,oreg}] appeared. Our results are
in agreement with theirs, where there is overlap.


\newpage

\centerline{\bf{Supplementary material }}

 
\section{Model for the Andreev bound state}

We consider a simple model of a superconducting double dot introduced by Leijnse {\it et al}\cite{leijnse}  to study the \lq{}poor man\rq{}s Majorana bound states\rq{},  coupled to a normal lead with the Hamiltonian given by
\beq
  H = H_{L} + H_{D} + H_{T} \eeq
\bea
 {\rm with} \quad 
  H_{L} &=& \sum_k\epsilon_{k} c^{\dagger}_{k} c_{k} \nonumber \\
  H_{D} &=& e_{1}d^{\dagger}_{1} d_{1} + e_{2}d^{\dagger}_{2} d_{2} + \Delta (d^{\dagger}_{1} d^{\dagger}_{2} + d_{2} d_{1}) \nonumber \\
 {\rm and} \quad  H_{T} &=& \sum_k c^{\dagger}_{k} (t^{\ast}_{1} d_{1} + t^{\ast}_{2} d_{2}) +h.c. \nonumber \\
 \eea
Here, $d_{1,2}$ are the dot degrees of freedom and $c_k$ are the lead fermions. The couplings of the lead to the two dots
are denoted by $t_{1,2}$. $e_{1,2}$ are the energies of the quantum dot levels in the two dots respectively ( we assume that the
dot is represented by a single level and $\Delta$ is the pair potential induced on the dots by proximity to a common superconductor.
 
The Hamiltonian for the dot  can be rewritten  as 
 $ H_{D} = e_{2} + \psi^{\dagger} h_{D} \psi$ with 
  $\psi = (d_{1}, d^{\dagger}_{2})$ and   
\bea
  h_{D} &= \begin{pmatrix}
        e_{1} & \Delta \\
        \Delta & -e_{2} \\
       \end{pmatrix}~. \nonumber \\
\eea
We can now diagonalise this Hamiltonian which gives
\beq
H_{D} = e_{2} + \tilde{\psi}^{\dagger} D \tilde{\psi} = e_{2} + \lambda_{+}a^{\dagger}a + \lambda_{-}b^{\dagger}b 
\eeq
wih diagonal eigenvalues 
\beq
\lambda_{\pm} = \frac{1}{2}[e_{1}-e_{2} \pm \sqrt{(e_{1}+e_{2})^{2} + 4\Delta^{2}}]
\eeq
and the diagonalising matrix (defined by $D= U^{\dagger}h_{D}U$)   
\beq
  U = \begin{pmatrix}
        \cos\alpha & \sin\alpha \\
        \sin\alpha & -\cos\alpha \\
       \end{pmatrix} ~.
\eeq
To obtain a zero energy state, we can choose 
$\lambda_{+} = e_{1} + \frac{\Delta^{2}}{e_{1}}$, 
$\lambda_{-} = 0 $  and the elements of the unitary rotation matrix to be of the form $\cos\alpha = \frac{e_{1}}{\sqrt{e_{1}^{2} + \Delta^{2}}}$ and 
$\sin\alpha = \frac{\Delta}{\sqrt{e_{1}^{2} + \Delta^{2}}}$. 

 Using the diagonalizing matrix $U$, the tunneling Hamiltonian can be rewritten in terms of the Bogoliubov operators $a, b$ as:
 \bea
  H_{T} &=& a^{\dagger}\underset{k} \sum [t_{1} \cos\alpha c_{k} - t^{\ast}_{2} \sin\alpha c^{\dagger}_{k}] \nonumber \\
        &+& b^{\dagger}\underset{k} \sum [t_{1} \sin\alpha c_{k} + t^{\ast}_{2} \cos\alpha c^{\dagger}_{k}] + h.c.
\eea
and projecting on to  the zero-energy subspace spanned by the operators $b$, $b^{\dagger}$ under the constraints $e_{2} = -\frac{\Delta^{2}}{e_{1}}$, $e_{1}>0$,
the tunnelling Hamiltonian becomes:
\begin{align}
  H_{T} &= b^{\dagger}\underset{k} \sum [u c_{k} + v^{\ast} c^{\dagger}_{k}] + h.c. \nonumber \\
  u &= t_{1} \sin\alpha =  \frac{t_{1}\Delta}{\sqrt{e_{1}^{2} + \Delta^{2}}} \nonumber \\
  v &= t_{2} \cos\alpha = \frac{t_{2}e_{1}}{\sqrt{e_{1}^{2} + \Delta^{2}}} \nonumber \\
  \end{align}
 which is precisely of the form given in the main text in Eq.14 for tunneling into an accidental zero-energy Andreev bound state.

\section{Derivation of the generalised Weidenmuller formula}
We consider the Hamiltonian of $N$ normal leads coupled to each other via a resonant  level (which we take here to be an ABS)
as well as via direct coupling,  discussed in the
main letter and given in Eqs. 5 - 8.

 To  derive the $S$ -matrix, we use the  equation of motion (EOM) method\cite{aleiner,hackenbroich} and write 
\begin{eqnarray}
 i\partial_{t} b(t) &=& [b,H](t) \nonumber \\
                       &=&  \underset{\alpha}{\sum}( t_{\alpha} \psi_{\alpha}(0,t) + v_{\alpha}^{\ast} \psi^{\dagger}_{\alpha}(0,t) ) \nonumber \\
 i\partial_{t}\psi_{\alpha}(x,t) &=& [\psi_{\alpha}(x),H](t) \nonumber \\
                       &=& -i v_{F} \partial_{x} \psi_{\alpha}(x,t) + \delta(x) (t_{\alpha}^{\ast} b(t) - v_{\alpha}^{\ast} b^{\dagger}(t))\nonumber \\
                       &&+ \delta(x) \underset{\beta}{\sum} (\tau_{\alpha \beta} + \tau_{\alpha \beta}^{\ast})\psi_{\beta}(0,t)
\end{eqnarray}
In terms of 
\begin{eqnarray}
t  &=& (t(1), t(2) \dots t(N))^T,  \nonumber \\
v &=& (v(1), v(2) \dots v(N))^T,  \nonumber \\
W &=&   \begin{pmatrix}
      t^{T} & v^{\dagger}\\
      -v^{T} & -t^{\dagger}\\
     \end{pmatrix}\nonumber\\ {\rm and} \quad 
T &=& \begin{pmatrix}
        (\tau+\tau^{\dag}) & 0 \\
                 0   &  -(\tau+\tau^{\dag})^{\ast}\\
       \end{pmatrix}
       \end{eqnarray} 
 and writing the wave-functions in the particle-hole basis as  
 \begin{eqnarray}      
 \psi(0,E) &=& (\psi_{P}(0,E), \psi_{H}(0,E))^T  \nonumber \\  &=& 
( \psi_{1}(0,E), \psi_{2}(0,E) \dots  \psi_{N}(0,E),  \nonumber \\ && \psi^{\dagger}_{1}(0,E), \psi^{\dagger}_{2}(0,E),\dots, \psi^{\dagger}_{N}(0,E))^T, 
\end{eqnarray}
we obtain
\begin{eqnarray}
 \psi(0+,E) &=& S(E) \psi(0-,E)  \nonumber \\ {\rm with} \quad 
 S(E) &=& [E(1 + i \pi \nu T) + i \pi \nu W^{\dagger}W]^{-1}  \nonumber \\
 &&[E(1 - i \pi \nu T) - i \pi \nu W^{\dagger}W] \nonumber 
\end{eqnarray}
Note here that  $\nu = {1}/{2 \pi v_{F}}$.
We now make contact with the usual Weidenmuller\cite{weidenmuller} formula by rewriting the scattering matrix as 
  \begin{eqnarray}
 S(E) &=& S_{N} - i \pi \nu (1 + i \pi \nu T)^{-1} W^{\dagger} \\ && [E + i \pi \nu W (1 + i \pi \nu T)^{-1} W^{\dagger}]^{-1}  
 W (1 + S_{N}) \nonumber \\
 {\rm where} \nonumber  \\
  S_{N} &=& (1 + i \pi \nu T)^{-1}(1 - i \pi \nu T) 
 \end{eqnarray}
so that in the absence of the direct tunneling term between the wires given by the $T$ matrix, the $S$-matrix reduces to the usual Weidenmuller formula - i.e.,
\begin{eqnarray}
 S(E) =1 - 2 i \pi \nu W^{\dagger} [E + i \pi \nu W W^{\dagger}]^{-1} W  
 \end{eqnarray}
 On the other hand, when there is no coupling to the resonant level, the $S$-matrix is just the usual tunneling matrix $T$ written in the particle-hole basis -
 \begin{eqnarray}
 S(E) = S_{N} \text{ for } W = 0
 \end{eqnarray}
  For tunneling to a MBS, we replace $b$ by the Majorana operator $\gamma$. Furthermore, the matrix $W$ is replaced by $W = (u,u^*)^T$.

\vspace{1.0cm}

\section{Expressions for finite bias subgap current and noise}
The expressions for the average current and the zero-frequency noise at normal leads for a junction of 
multiple normal leads connected to a grounded superconductor are given by\cite{anantram}
\begin{eqnarray}
 I_{p} &=& \frac{e}{h} \underset{k \in \{N,S\},\alpha\gamma} \sum sgn(\alpha) \int_{0}^{\infty} dE A_{k\gamma,k\gamma}(p,\alpha,E) \nonumber \\
       &&f_{k\gamma}(E)  \nonumber \\
 S_{pq} &=&  \frac{e^{2}}{h} \underset{k,l \in \{N,S\},\alpha\beta\gamma\delta} \sum sgn(\alpha)sgn(\beta) \nonumber \\
 &&\int_{0}^{\infty} dE A_{k\gamma,l\delta}(p,\alpha,E) A_{l\delta,k\gamma}(q,\beta,E) \nonumber \\
 &&f_{k\gamma}(E) (1-f_{l\delta}(E)) \nonumber \\
 A_{k\gamma,l\delta}(p,\alpha,E) &=&  \delta_{pk} \delta_{pl} \delta_{\alpha \gamma} \delta_{\alpha \delta}- (s^{\alpha\gamma}_{pk})^{\ast}(E) s^{\alpha\delta}_{pl}(E) \nonumber \\
 f_{k\gamma}(E) &=& [1 + e^{\frac{E-sgn(\gamma)\mu_{k}}{k_{B}T}}]^{-1} \nonumber \\
 sgn(\alpha) &=&  \begin{cases} 
      1 & \alpha = e \\
     -1 & \alpha = h
   \end{cases}
 \end{eqnarray}
These were the equations used to obtain the figures in Figs. 2 and 3 of the main text.

%
%
%

\end{document}